# Local search on protein residue networks

Susan Khor
slc.khor@gmail.com
June 9, 2013

**Abstract**

Self-avoiding random walks were performed on protein residue networks. Compared with protein residue networks with randomized links, the probability of a walk being successful is lower and the length of successful walks shorter in (non-randomized) protein residue networks. Fewer successful walks and shorter successful walks point to higher communication specificity between protein residues, a conceivably favourable attribute for proteins to have. The use of random walks instead of shortest paths also produced lower node centrality, lower edge betweeness and lower edge load for (non-randomized) protein residue networks than in their respective randomized counterparts. The implications of these properties for protein residue networks are discussed in terms of communication congestion and network vulnerability. The randomized protein residue networks have lower network clustering than the (non-randomized) protein residue networks. Hence, our findings also shed light on a hitherto neglected aspect: the importance of high network clustering in protein residue networks. High clustering increases navigability of a network for local search and the combination of a local search process on a highly clustered small-world network topology such as protein residue networks reduces communication congestion and network vulnerability.

## 1. Introduction

Protein residue networks represent protein macro-molecules as graphs. The nodes of a protein residue network (PRN) represent the amino acid molecules of a protein and the edges denote node pairs which are within some Euclidean distance from each other. PRNs are small-world networks, i.e. they are heavily clustered at the local level, but have inter-nodal ties that effectuate short diameter and average path lengths at the global level [1]. Short average path length is believed to be an important topological feature to facilitate interaction cooperativity crucial for rapid and correct protein folding [2, 3, 4]. Less has been said however about the role of high clustering in PRNs.

The small-world property of protein residue networks was previously established with the shortest path algorithm which is a global search. However, the average path length (geodesic or chemical distance) of a network is not always a determining factor in network communications [5, 6]. In this research, we investigate decentralized search on PRNs by taking self-avoiding random walks (SARWs) on them. SARWs were previously shown to be the most effective local search strategy on several different network topologies [7]. SARWs were used by Kleinberg to demonstrate that not all small-worlds are equal from a local search perspective (we note that his result does not depend on the condition of self-avoidance) [8]. Kleinberg's result highlights the importance of contextual clustering, i.e. the existence of meaningful local structure, in the navigability of small-worlds. Since we are motivated in part to understand why PRNs have a high level of clustering, the use of SARWs is a reasonable fit.

## 2. Method

### 2.1 Network models and characteristics

The protein residue networks (PRNs) were constructed using the coordinates of the Cα atom of amino acids from the Protein Data Bank [9]. Define PCM as the adjacency matrix of a PRN. Since PRNs are undirected and in our case unweighted, PCMs are symmetric and binary. PCM $(i, j) = 1$ if the node pair $(i, j)$ is situated less than 7Å from each other, otherwise PCM $(i, j) = 0$. An edge or node pair $(i, j)$ is considered long-range if $| i - j | > 9$. Long-range edges connect amino acids which are distant in the primary structure but are in close spatial proximity in the tertiary structure [10]. Fig. 1 summarizes some basic statistics about the eight PRNs in our dataset. The number of links or edges increases linearly with the number of nodes [11], and a linear relationship between number of links and number of nodes holds even when the edges are split into short-range (SE) and long-range (LE). These linear relationships are typical for PRNs [12].

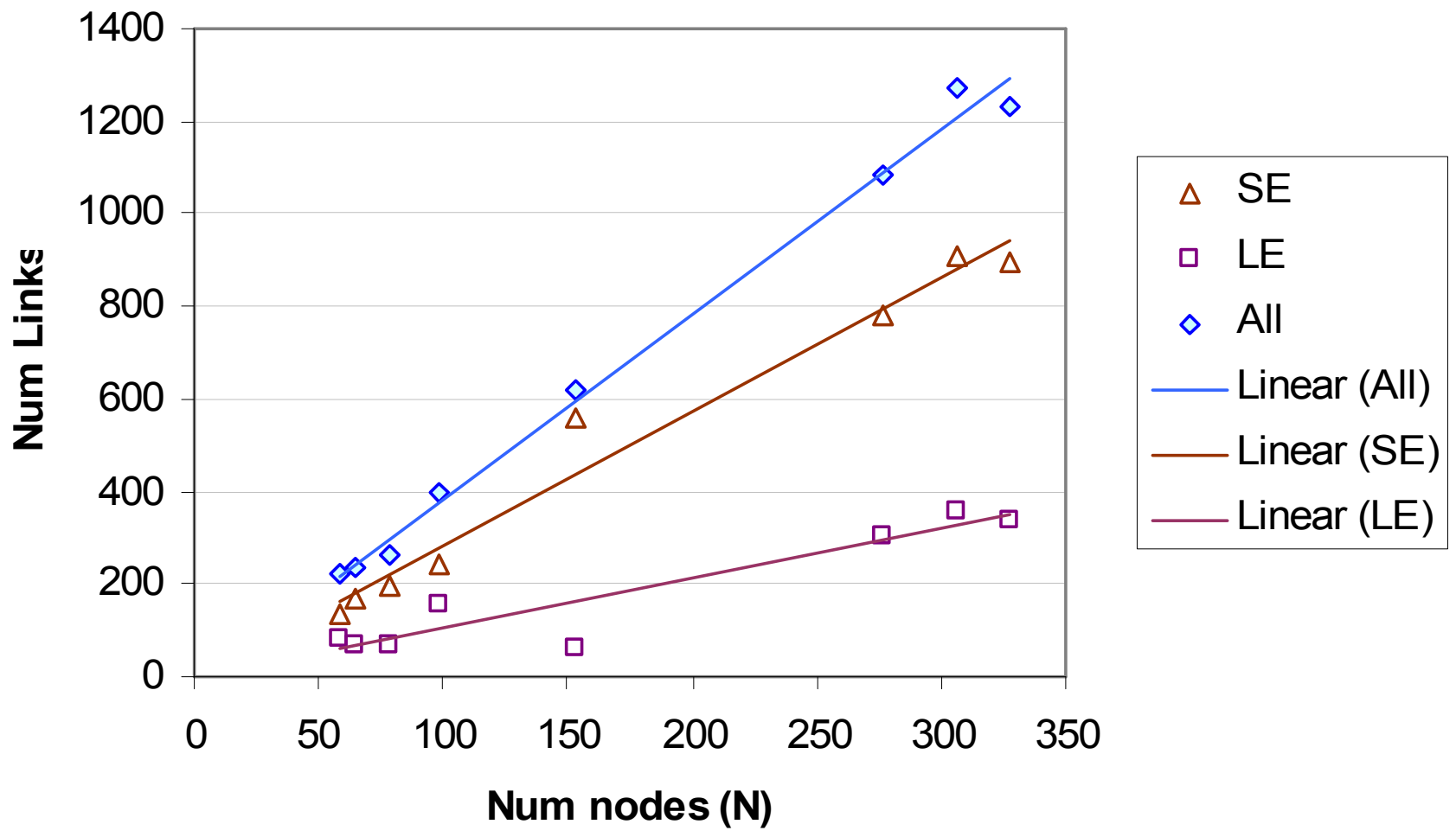

Fig. 1 Number of nodes and links of the eight protein residue networks (PRNs) in this study. The number of links or edges increases linearly with the number of nodes [11] even when the set of links are partitioned in short-range (SE) and long-range(LE) sets [12]. An edge $(i, j)$ is long-range if $| i - j | > 9$ [10].

Randomization of PRNs is achieved in two ways: (i) by rewiring the short-range edges and keeping the long-range edges intact, and (ii) by rewiring the long-range edges and keeping the short-range edges intact. Randomized PRNs of the first kind are labeled randSE and of the second kind randLE. Edge rewiring is performed such that the number of links and the node degrees do not change [13]. This can be done by picking two edges at random from the variable set of edges e.g. e1(a, b) and e2(c, d), exchanging one of their endpoints e.g. e3(a, c) and e4(b, d) and replacing e1 and e2 with e3 and e4 only if e3 and e4 do not already exist in the set of all currently existing edges. 20 randSE and 20 randLE networks were generated for each of the eight PRNs in our dataset. All reported statistics for randSE and for randLE are averaged over 20 networks.

Fig. 2-left compares the length of edges in non-randomized and randomized PRNs. The length of an edge $e$ with end points at $i$ and $j$ is the sequence distance between $i$ and $j$, or SD($e$) = | $i - j$ |. randSE networks have significantly longer edges on average than randLE networks and the (non-randomized) PRNs. This is partly due to there being more short-range edges than long-range edges (Fig. 1) and that the short-range links involve all the nodes in a PRN [12]. Therefore there is more mixing in randSE networks and they become more like random graphs.

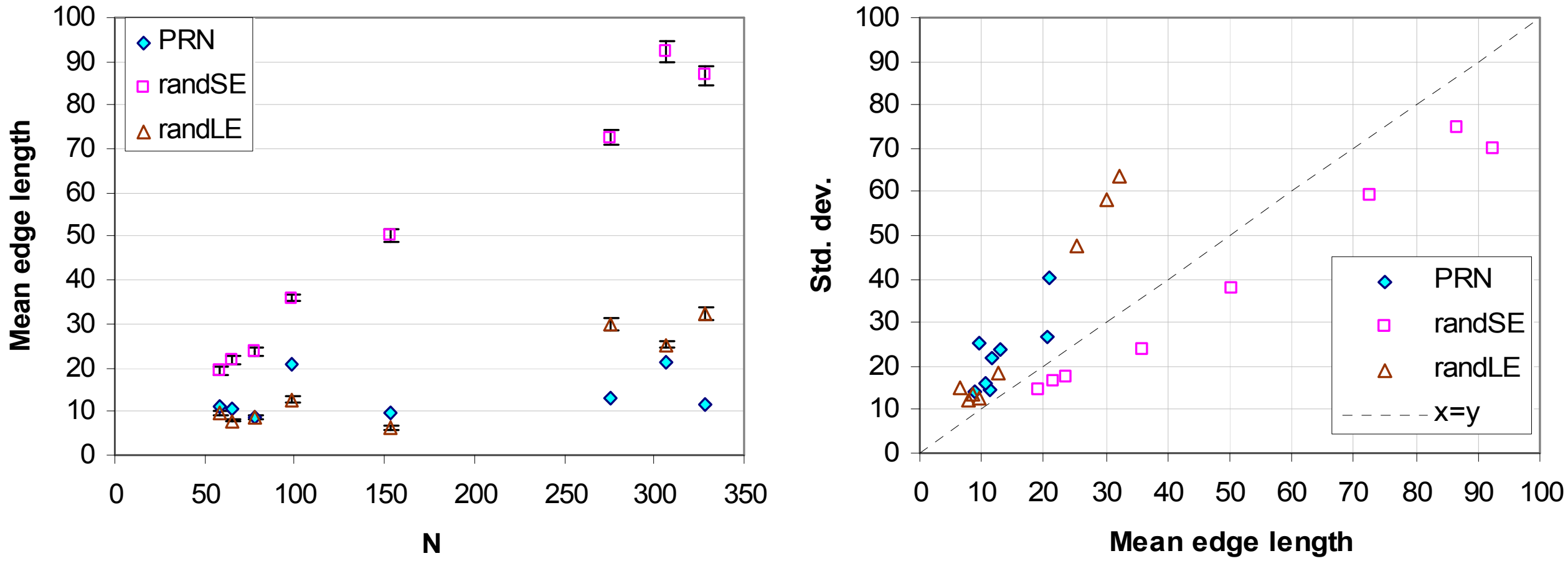

Fig. 2 Effect of randomization on edge length. Left: On average, edges are significantly longer in randSE networks. The length of an edge is the sequence distance between its two endpoints. Error bars mark one standard deviation from the mean. Right: The distribution of edge lengths in randSE networks show less variance.

The distribution of edge lengths in randSE networks is distinct from the edge length distribution of randLE networks and PRNs (Fig. 2-right). randSE networks have a coefficient of variation (standard deviation to mean ratio) which is less than 1, while the coefficients of variation for randLE networks and PRNs are more than 1. While the edge length distributions of PRNs were right skewed, we did not observe a power-law tail to their edge length distributions. This could be a point for further study. The significance of a network's edge length distribution has been studied for networks embedded in physical space [e.g.: 14-18] and found to influence a network's dynamics [e.g.: 8 & 19]. Several of these studies use theoretical models which bias against long-range links. A common model is to define the probability of creating a link of length λ as $p(\lambda) = \lambda^{-\alpha}$.

Small-world networks are characterized by two network properties: (i) high clustering levels and (ii) average path length which increases logarithmically with the number of nodes in a network [1]. Define the clustering coefficient of node $x$ as CLUS($x$) = $\frac{2\, e_x}{k_x (k_x - 1)}$ where $k_x$ is the degree of node $x$, and $e_x$ is the number of links that exist amongst the $k_x$ nodes. The clustering coefficient of a network G with N nodes is CLUS(G) = $\frac{1}{N} \sum_i^N CLUS(i)$. Define SPL($x$, $y$) as the length of a shortest path (in graph distance) between nodes $x$ and $y$. We assume that SPL($x$, $y$) = SPL($y$, $x$) to reduce computation time. The average path length of a network G with N nodes is APL(G) = $\frac{2}{N(N-1)}\sum_{i<j}^{N} SPL(i,j)$ . SPL($x$, $y$) = 0 if no path exists between $x$ and $y$. By this definition, both PRNs and randLE networks are small-world networks, and randSE networks are not (Fig. 3-top). The effects of randomization were previously explored in greater detail in [12].

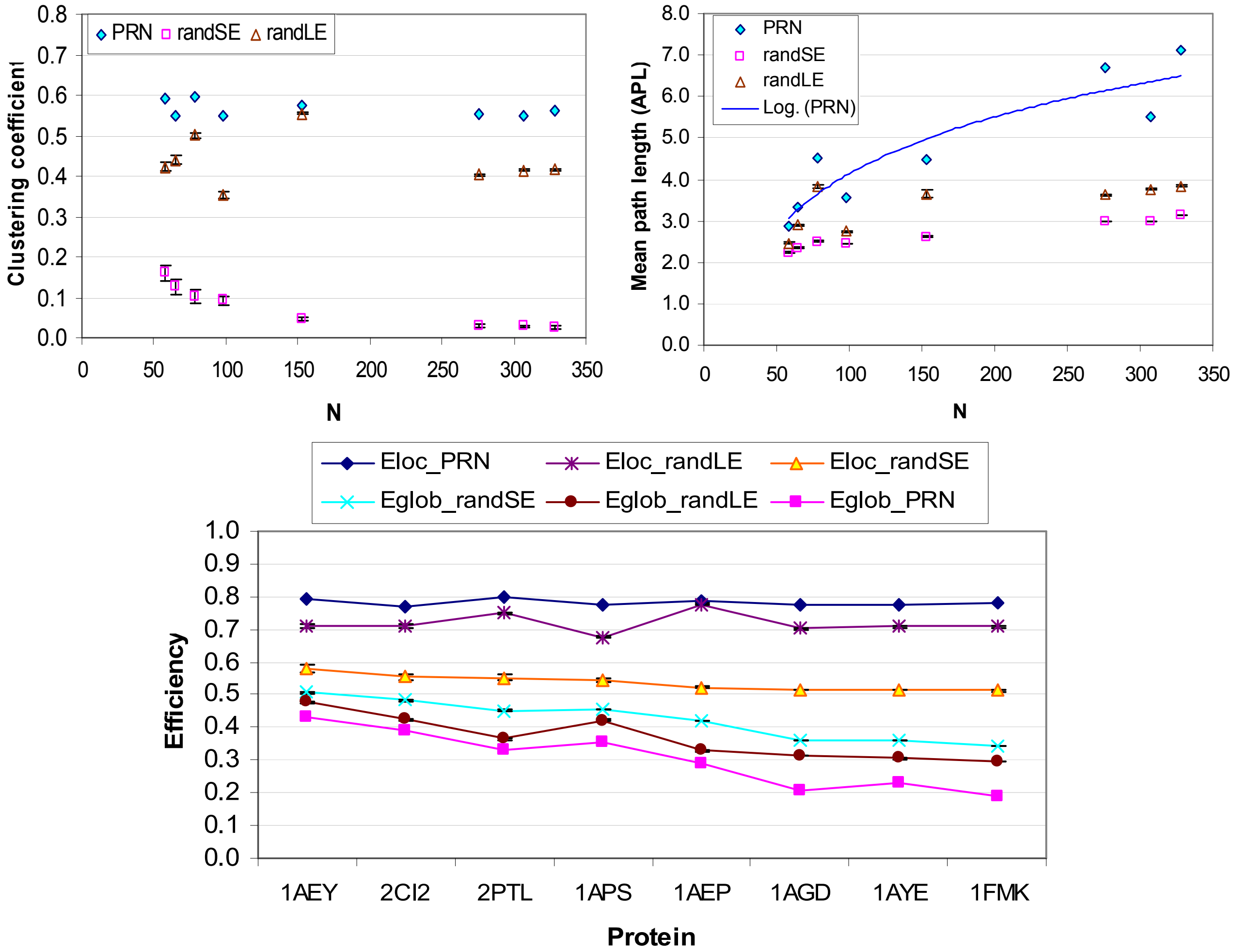

Fig. 3 Effect of randomization on the small-world network property of PRNs. Both PRNs and randLE networks remain as small-world networks, while randSE networks become more like classical (Erdos-Renyi) random graphs. Error bars (often with an interval too small to be seen) denote one standard deviation from the mean. Top: Randomization of short-range links destroys local structure in PRNs, and the clustering coefficient of randSE networks are significantly smaller (top-left). However or as a result, randSE networks enjoy significantly shorter path lengths on average. The average path length of PRNs scales logarithmically with N (top-right). Bottom: From a communication efficiency view point, randSE networks are most efficient globally (due to their shorter average path lengths), and least efficient locally (due to their weaker clustering). In contrast, PRNs are least efficient globally and most efficient locally.

An alternative way of characterizing small-world networks was proposed in [20] and it is expressed in terms of efficiency of information exchange at the global and local levels. Small-world networks are both globally and locally efficient. The global efficiency of a network G is defined as Eglob(G) = $\frac{2}{N(N-1)}\sum_{i<j}^{N}\frac{1}{SPL(i,j)}$. Here, $\frac{1}{SPL(x,y)}=0$ if no path exists between $x$ and $y$. By definition, the presence of longer shortest paths will reduce global efficiency and a complete graph has maximum global efficiency with Eglob = 1.0. The local efficiency of a network G is defined as Eloc(G) = $\frac{1}{N}\sum_{i}^{N}Eglob(G_i)$ where $G_i$ is a subgraph of G comprising node *i* and its direct neighbors. In a highly clustered network, many of its subgraphs will be globally efficient since it is likely that the direct neighbors of a node are linked to each other. Nodes directly linked to each other have maximum communication efficiency $\frac{1}{SPL(x,y)}=1$.

Fig. 3-bottom compares the global and local efficiencies of PRNs, randSE and randLE networks. Of the three types of networks, PRNs have the lowest global efficiency (Eglob) values but the highest local efficiency (Eloc) values. Randomization reduces Eloc (reduces clustering), and increases Eglob (reduces average path length). This effect is more pronounced when short-range edges are randomized. This result is expected given that the short-range links of a PRN are more highly clustered than the long-range links [11, 12]. Due to their highly clustered short-range links, randLE networks remain small-world networks [12].

## 2.2 Local search algorithm: self-avoiding random walk (SARW)

The local search is conducted via self-avoiding random walks. The search is local since at each step, only information local to the current node, that is knowledge of its direct neighbors and the nodes visited so far, is used to determine (by a uniform random selection) the next node in the walk. Note that our implementation inspects all direct neighbors of a node in random order until a suitable (not yet visited by the walk) next node is found before concluding that there is no suitable next node. A SARW terminates when it reaches its target node or when it cannot make further moves (for instance when it finds itself in a loop). We assume that the network is congestion free.

A SARW that reaches its target node is a successful one. The length of a SARW (walk length) is the number of edges in the path that it traversed, or one less than the number of nodes it visited. For the graph in Fig. 4, four SARWs starting from node *s* with node *t* as the target are equally possible. All except one SARW reached the target node *t*. A SARW may or may not be a shortest path.

SARW between all pairs of nodes were performed on each network. In total, we have 328 networks (8 PRNs, 8×20 randSE networks and 8×20 randLE networks). Due to the random nature of the walks, we repeated the local search 20 times for each PRN; this gives a total of 480 walks. Walk statistics for a PRN or its randomized counterpart, are calculated over the corresponding 20 walks.

| SARW | Success | Walk length |
|---|---|---|
| s - t | Y | 1 |
| s - a - t | Y | 2 |
| s - a - b | N | 2 |
| s - b - a - t | Y | 3 |

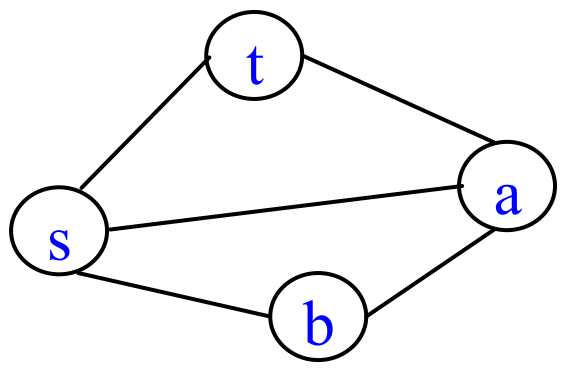

Fig. 4 The table on the left lists the four possible self-avoiding random walks starting at *s* with target node *t*. Of these, only three successfully reach *t*. The shortest SARW is of length 1, while the longest SARW took three steps before reaching *t*. The s-a-b walk stopped before reaching *t* and therefore failed to reach *t* because it could no longer take a step from *b* without revisiting either *s* or *a*.

### 3. Results: Walk length and local search success

SARWs between all node pairs on PRNs were significantly shorter than SARWs on randSE networks (Fig. 5). This is so in spite of PRNs having longer APLs (Fig. 3 top-right). As N increases, walk lengths on randSE networks scale linearly, while the increase is only logarithmic on the small-world networks, i.e. PRNs and randLE networks.

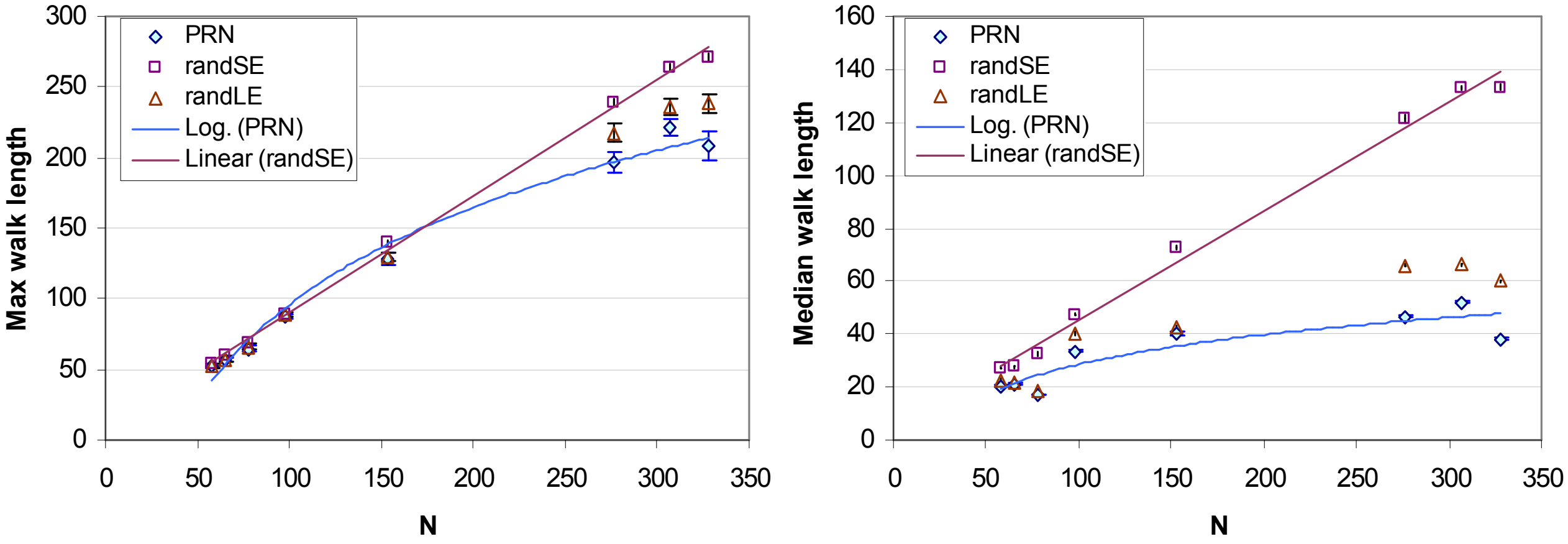

Fig. 5 The average maximum (left) and median (right) length of SARWs between all node pairs in a network. Error bars (often with an interval too small to be seen) denote one standard deviation from the mean computed over 20 corresponding SARWs. The median walk length is reported because of the high variance in walk lengths per network. The signal-to-noise (mean to standard deviation) ratio of SARWs is greater than 1.0 but less than 2.0; the signal-to-noise ratio for shortest paths in Fig. 3 is more than 2.0.

The SARWs on randSE networks were more successful, i.e. a larger proportion of the SARWs on randSE networks reached their respective target nodes (Fig. 6 top-left). Hence compared with randSE networks, SARWs between all node pairs on PRNs were less successful, but were shorter when successful (Fig. 6 middle- and bottom-left).

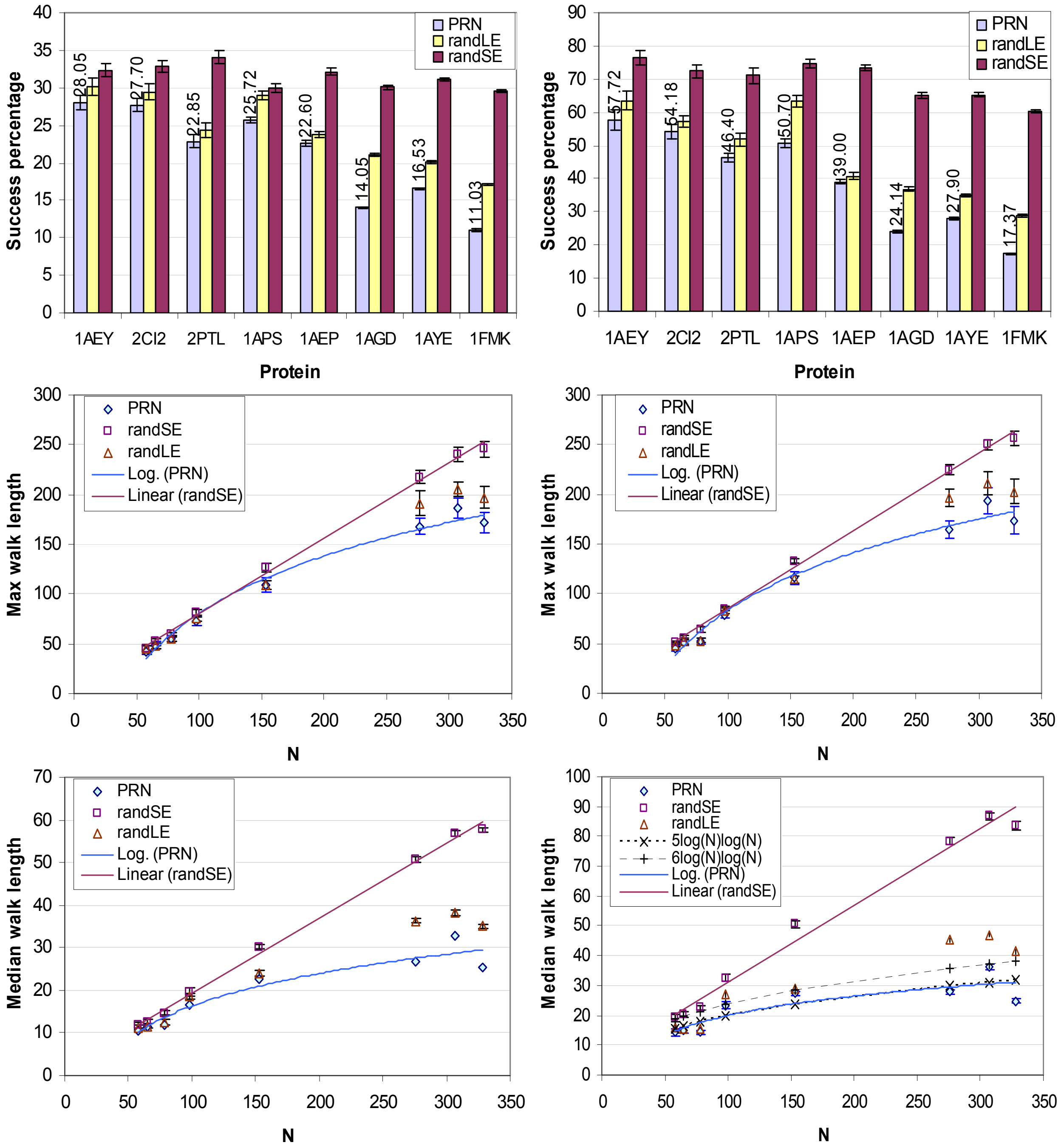

Fig. 6 Effect of randomization on local search success and length of successful walks. On the left, the SARWs between all nodes are evaluated. On the right, only the SARWs between nodes with above average closeness centrality are considered. The effects of randomization are the same in both cases. A larger proportion of the SARWs on randSE networks reached their respective target nodes, but took longer to do so. As the number of nodes (N) increases, the length of walks in the poorly clustered randSE networks increase linearly, while in the richly clustered PRNs and randLE networks, the increase in walk length is logarithmic. We submit that fewer successful but shorter walks promote specificity in inter-residue communication within protein molecules.

Higher levels of clustering in PRNs create loops which can trap SARWs and prevent them from reaching their respective destinations[1]. This behavior may be useful to absorb or localize the after effects of perturbations to maintain protein stability [4]. Paradoxically, higher levels of clustering in PRNs may also be responsible for shorter walks [8], thus enabling fast communication between residues for protein activity [4]. But how clustering helps to identify or create long-range links in the case of PRNs is still an open question. In social networks and other systems that exploit local structure [e.g. 21], there are latent cues or problem-specific indexes built to guide local search. What are these cues in proteins?

Producing fewer successful walks and shorter successful walks imply a level of specificity in inter-nodal communication. Previous studies have traced specific communication pathways within a protein molecule [4]. Further, protein sites are not all equal. Protein residues at certain sites are more actively involved in protein activity or more evolutionarily conserved than others, and a select few act as protein folding nucleation sites. It is the fast communication between these functional residues located at different regions of a protein which is crucial for protein folding and subsequent activity.

Ref. [22] observed a positive correlation between the functional significance of a site and the closeness of a site to all other sites in a protein. This finding was subsequently used in [23] to identify protein folding nucleation sites in PRNs. The closeness of a node (protein site) to all other nodes in a network is its closeness centrality defined as CLOSE($x$) = $\left. N-1 \middle/ \sum_{i \neq x}^{N} SPL(x,i) \right.$. Nodes with high closeness centrality are closer (in graph distance) to other nodes in the network. Following from their shorter APLs or higher global efficiency (Fig. 3), nodes in a randSE network have higher closeness centrality on average than nodes in a PRN (Fig. 7).

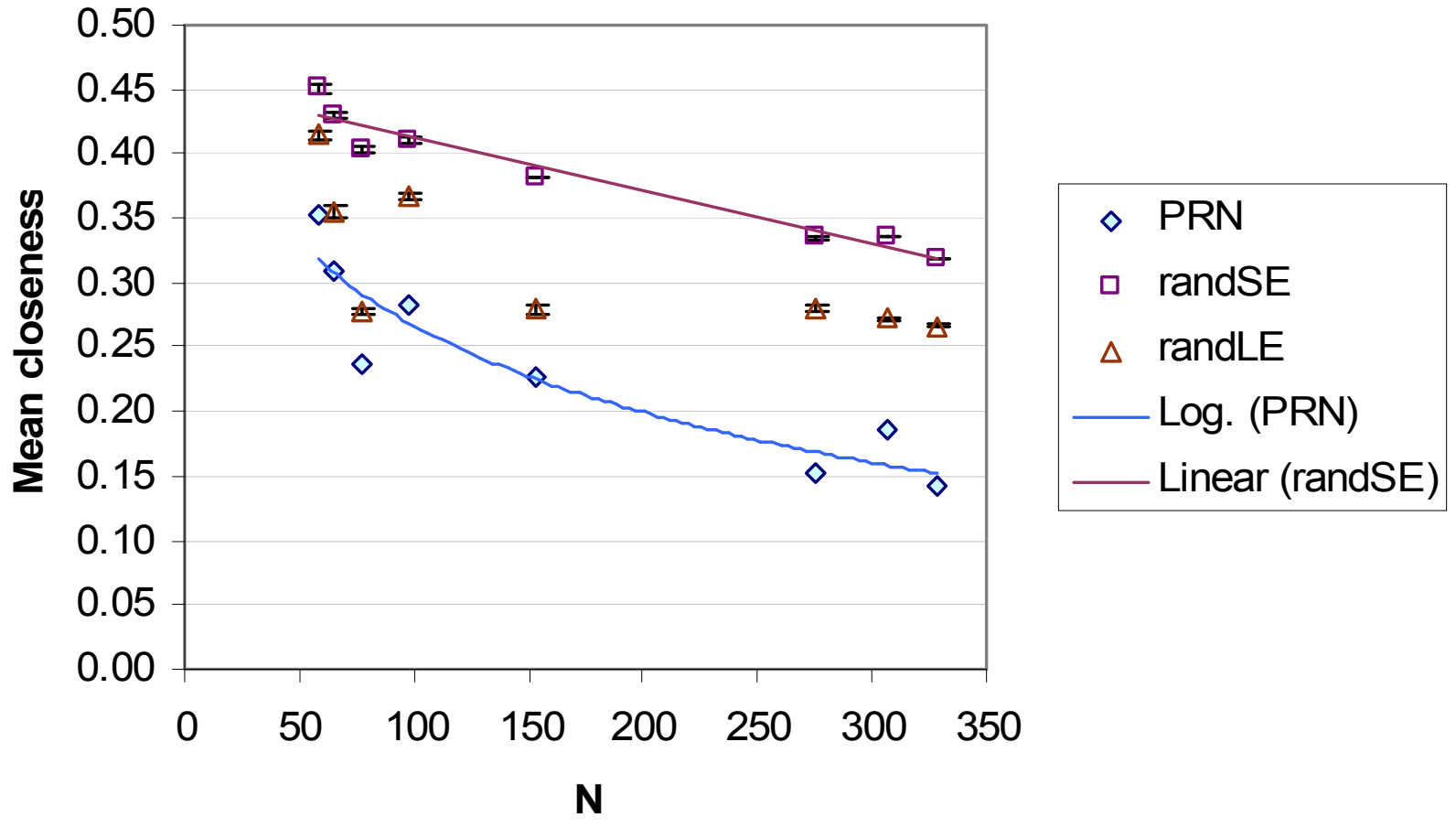

Fig. 7 Nodes in a PRN have lower closeness centrality on average, i.e. they are further away (in graph distance) from other nodes within in the same network. This follows from the longer APLs of PRNs (Fig. 3). Nonetheless, SARWs on PRNs are shorter than SARWs on randSE networks (Fig. 5).

[1] Adding multiple edges between already connected node pairs will give a random walk a limited number of chances to revisit a node. This kind of self-avoiding random walk would avoid revisiting edges instead of nodes.

If the SARWs are restricted to pairs of source and target nodes with higher closeness centrality than average (larger than the mean), we find again that PRNs produce fewer successful walks (Fig. 6 top-right) but shorter successful walks (Fig. 6 middle- and bottom-right). We also observe that the length of successful walks on PRNs grow logarithmically with network size (more precisely it is bounded by $c(\log N)^2$ which is in agreement with Kleinberg's model when $\alpha=2$ [8]); whilst there is a linear relationship between walk length and network size for randSE networks. So even though randSE nodes have higher closeness centrality on average than PRN nodes (Fig. 7), SARWs on PRNs are significantly shorter and lengthen at a slower pace with increases in network size. It could be interesting to redo the studies in [22 & 23] with random walks instead of shortest paths.

**4. Results: Node centrality**

Node centrality is a measure of a node's involvement in the dynamics of a network. It has been quantified in a number of ways: e.g. node degree, number of shortest paths passing through a node, and number of random walks passing through a node [24]. Ref. [2] observed that node centrality (using the shortest path method) changes as a protein folds towards its native state and that residues with high (large) node centrality during the transition state play a critical role in the folding process. Ourselves, we observed that variation in node centrality (using the shortest path method) between residues of a protein decreases as protein folding progresses in our simplified model of restoring edges to a PRN in monotonically non-decreasing sequence distance order.

The influence of node centrality (using the shortest path method) has also been studied in the context of network congestion. The key lesson is the reciprocal relationship between maximum node centrality and the congestion threshold, which is the critical rate of packet creation above which network congestion will occur [25]. Subsequent studies have formulated strategies to reduce congestion by adjusting network topology to reduce maximum node centrality [e.g. 26]. For proteins, the congestion threshold may be viewed as the maximum amount of stimulation (e.g. interaction with other proteins and molecules or the water substrate) a protein can process simultaneously without detriment to itself. Given the many interactions that a protein makes in a cell during its lifetime and adopting a Panglossian view of Nature, we expect PRNs to have lower (smaller) node centrality than their randomized counterparts.

However, we see in Fig. 8-left that this is not the case when shortest paths are used to compute node centrality. When SARWs are used to compute node centrality, we observe the reverse or expected situation (Fig. 8-right): the maximum and median node centrality values of PRNs are now significantly smaller on average than those for randSE networks. This reversal is a consequence of relative path or walk length. randSE networks have shorter path lengths than PRNs (Fig. 3), but longer walk lengths than PRNs (Fig. 5). Longer paths or walks increase the probability of node visitations and by definition node centrality.

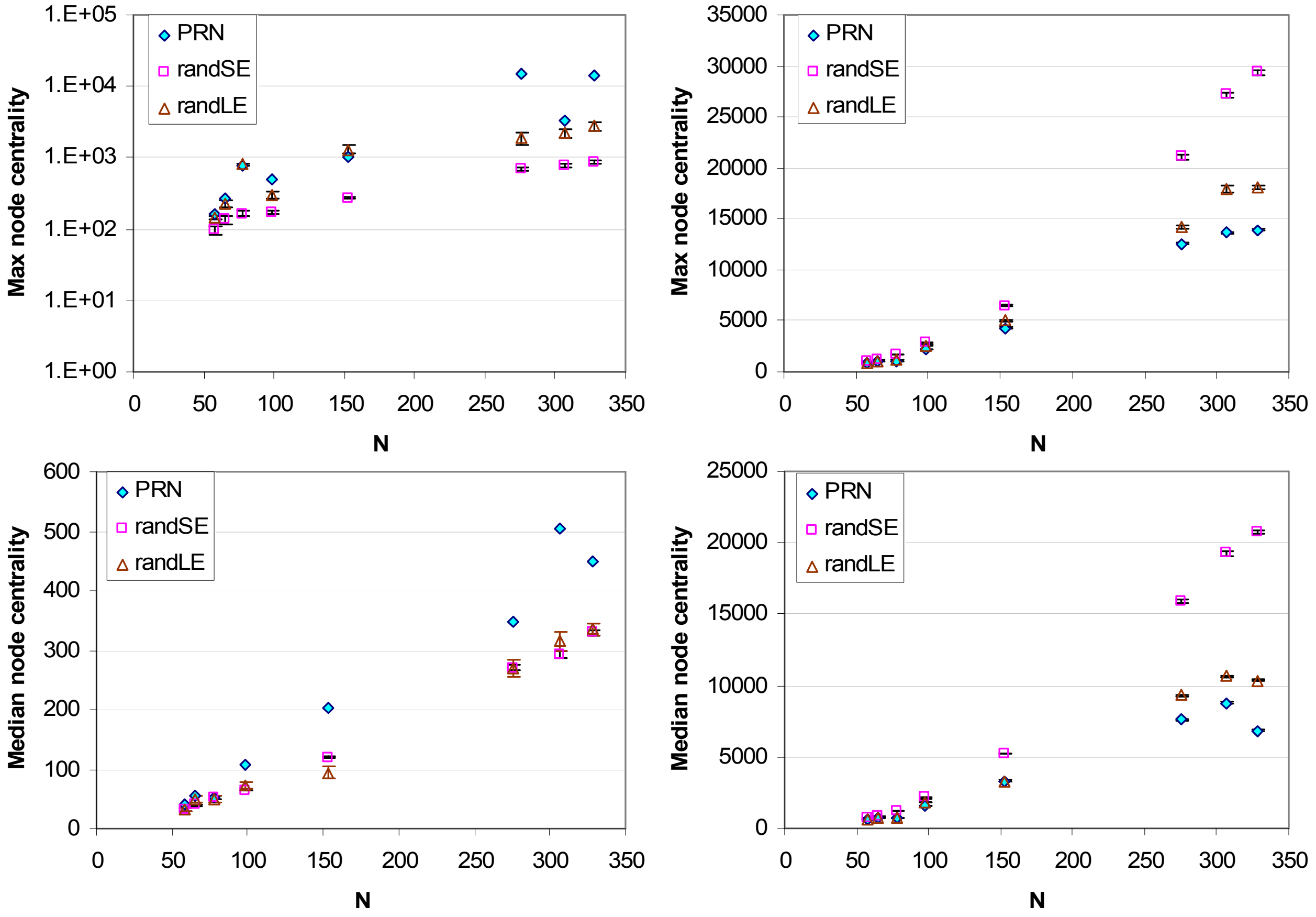

Fig. 8 Effect of computing node centrality with SARWs instead of shortest paths. On the left, shortest paths between all nodes are used to compute node centrality, and nodes in PRNs have significantly higher centrality on average than nodes in randSE networks. On the right, SARWs between all nodes are used, and nodes in randSE networks have significantly higher centrality on average than nodes in PRNs.

## 5. Results: Edge betweeness and edge load

Edges can also be examined in terms of their centrality or importance in network dynamics. Intuitively, an edge that is traversed in many paths, be they shortest paths or SARWs, is an important edge. Define the betweeness of an edge as the number of paths traversing the edge in a set of paths for a network.

The presence of edges with high betweeness increases a network's vulnerability since their disruption could potentially affect many pathways. Fig. 9 shows that there is more distinction between the three types of networks in terms of edge betweeness when the paths are SARWs than when they are shortest paths. PRNs exhibit lower (smaller) edge betweeness than randSE networks. This suggests to us a level of topological robustness in PRNs. Indeed, proteins are fairly robust to random attacks according to mutagenesis studies [22], and possess alternative pathways or redundant links between critical sites [4].

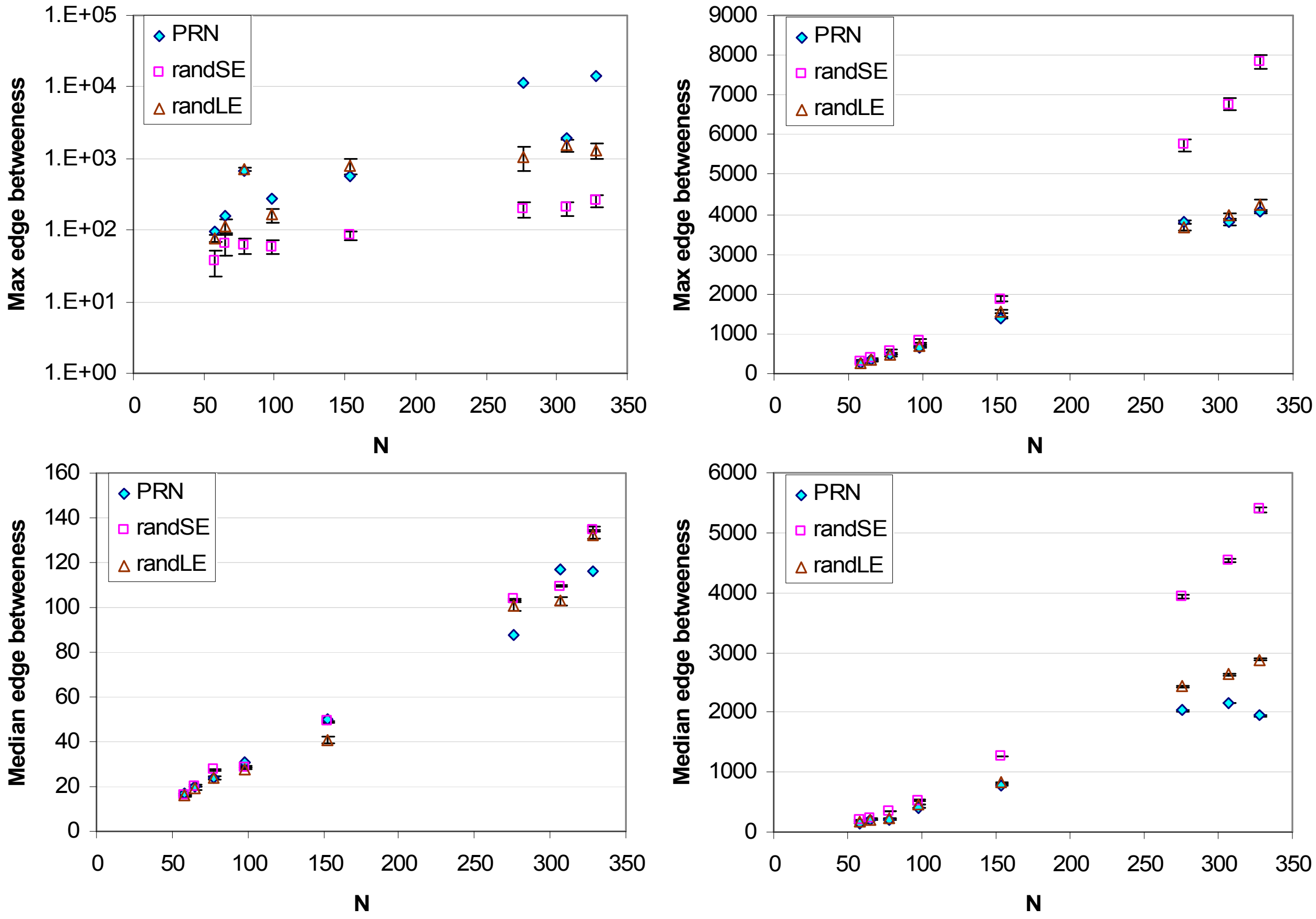

Fig. 9 Effect of computing edge betweeness with SARWs instead of shortest paths. On the left, shortest paths between all nodes are used to compute edge betweeness, and there is little difference between the three kinds of networks. On the right, SARWs between all nodes are used, and edges in randSE networks have significantly higher betweeness on average than nodes in PRNs. Higher edge betweeness in randSE networks implies uneven usage of edges in a network.

A measure similar to edge betweeness is used in the Connection Graph Stability method [5] to estimate coupling coefficients for complete synchronization. Instead of counting the number of paths that traverses an edge, the sum of the length of paths that traverses an edge is used to compute the load of an edge. A link with the maximum load is a weakest link, and the maximum load value is the critical value for all coupling strengths above which the synchronization becomes globally asymptotically stable. This critical value may be lowered by smart selection of paths that distributes edge loads as homogeneously as possible or by adding edges to create alternate paths. The creation of alternate paths with conservative increase in average distance is also a strategy employed to reduce network congestion [26]. Load balancing observed via the changes in node centrality and in edge betweeness as protein folding is simulated by the restoration of links to a PRN in monotonic increasing sequence distance order may be an important idea for solving the problem of evolving realistic protein contact maps.

As with node centrality (Fig. 8), when shortest paths are used to compute edge load PRNs exhibit heavier (larger) edge loads than randSE networks (Fig. 10 left). But when SARWs are used to compute edge load PRNs exhibit lighter (smaller) edge loads than randSE networks (Fig. 10 right). Lighter edge loads imply stronger weakest link(s) and easier synchronizability.

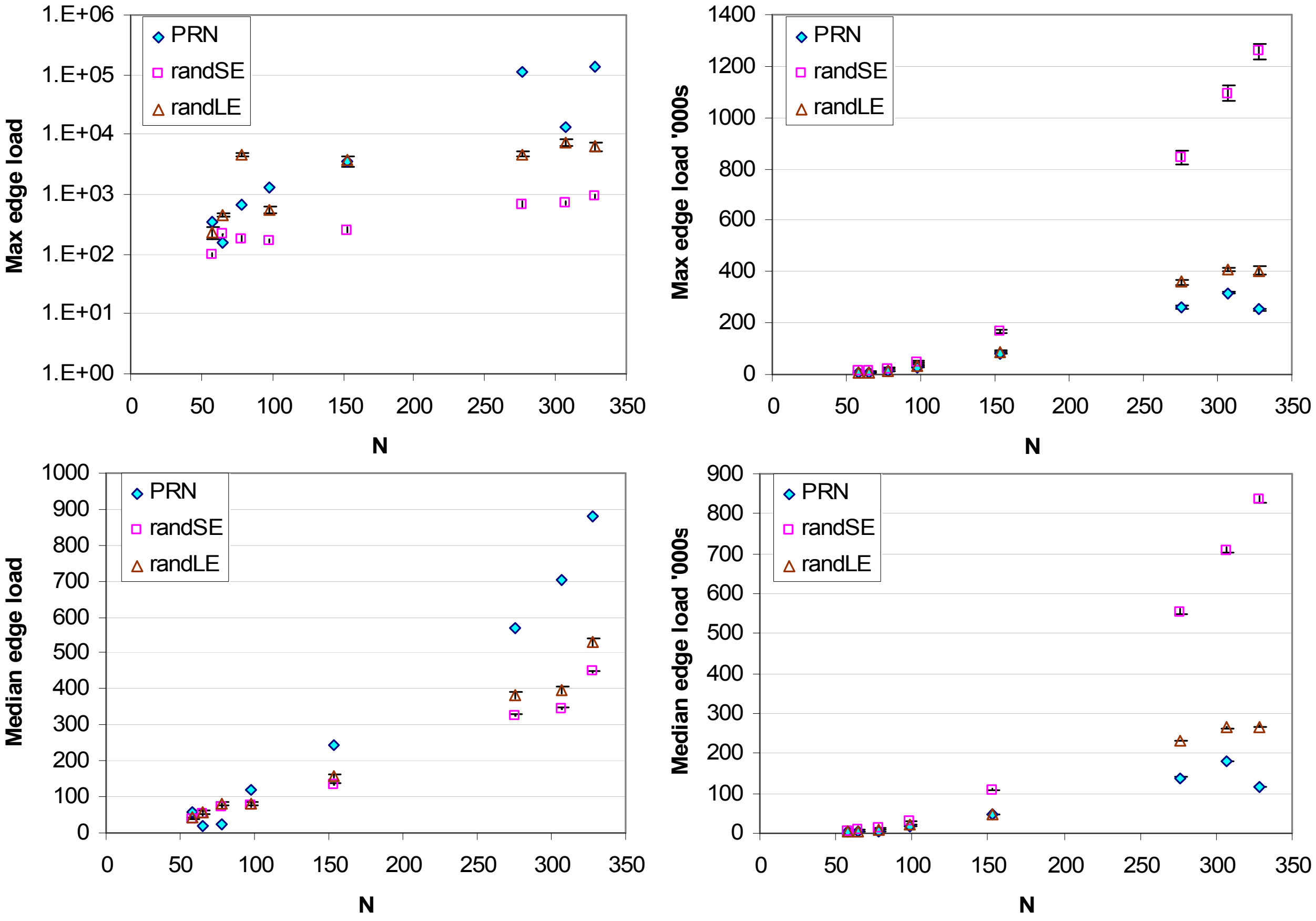

Fig. 10 Effect of computing edge load with SARWs instead of shortest paths. On the left, shortest paths between all nodes are used to compute edge load, and edges in PRNs have significantly higher edge load on average than edges in randSE networks. On the right, SARWs between all nodes are used, and edges in randSE networks have significantly higher edge load on average than nodes in PRNs.

## 6. Result: Mean path cost

Mean edge length (Fig. 2-left) can be viewed as the average cost of a link in a network. Such cost may include the effort involved in wiring a system, or the risk associated with the formation of a link. Define the mean cost of paths in a network as APC(G) = AL(G) × AEC(G). The average cost of an edge in a graph G with E edges is AEC(G) = $\frac{1}{E}\sum_{e}^{E} SD(e)$ where SD($e$) = $|\,i-j\,|$ is the sequence distance of the edge $e$ with endpoints at $i$ and $j$. AL(G) is the average length of paths in G. It may be the average length of shortest paths or the average length of SARWs. In either case, the average cost of a path in PRNs is significantly lower than the average cost of a path in randSE networks (Fig. 11). Thus it is inefficient in terms of wiring cost to do random wiring even when shortest paths are desired. The question whether PRNs are optimal in terms of wiring cost is interesting but not explored here.

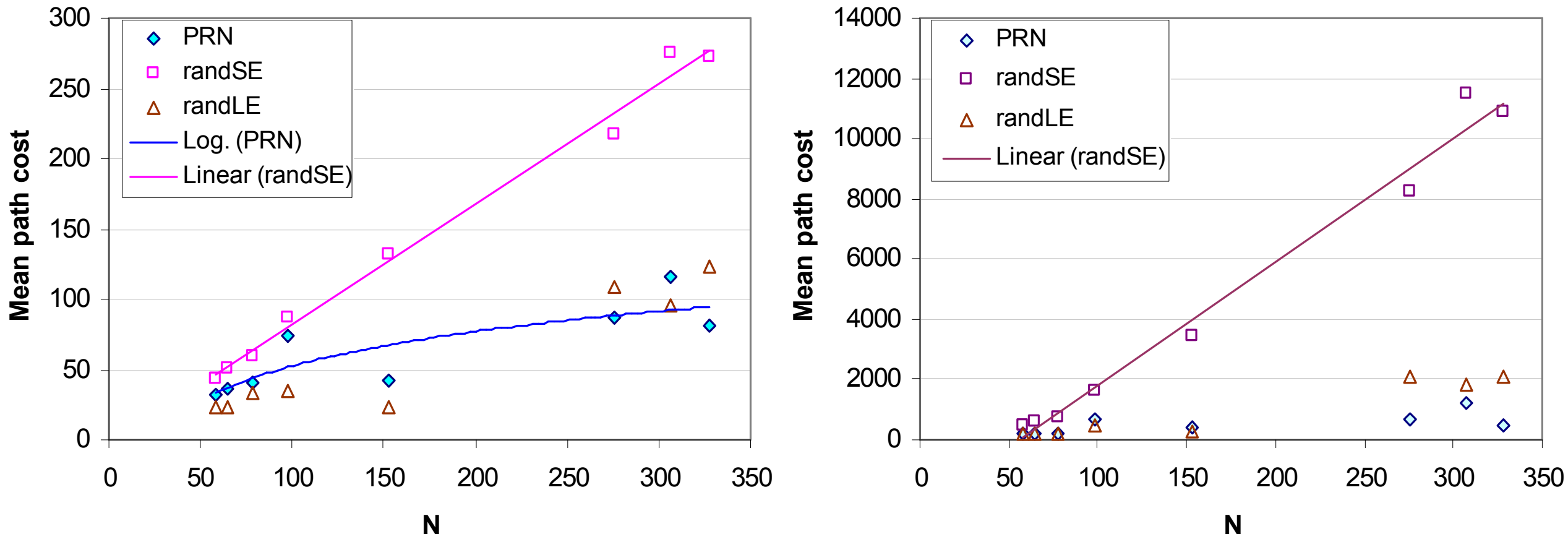

Fig. 11 PRNs have significantly lower mean path cost than randSE networks. Left: Mean path cost is computed with average length of shortest paths. Right: Mean path cost is computed with average length of SARWs.

## 7. Concluding remarks

Protein residue networks (PRNs) are small-world networks. The small-world property of protein residue networks was previously established with a global search for a collection of shortest paths between all node pairs. We take a different approach and investigate local search on protein residue networks. Our decentralized search produces a collection of self-avoiding random walks (SARWs) starting from each node in the network and terminating either at the assigned destination node or at the node where a walk could no longer proceed without revisiting a node.

We found that PRNs exhibit network properties that we argue are favourable for protein folding and protein activity when SARWs are used to investigate the dynamics on PRNs than when shortest paths are used. With SARWs, information transmission on PRNs appear more specific but still efficient (section 3), less prone to congestion and disruption (sections 4 & 5), easier to synchronize (section 5) and more economical (section 6). This persuades us that paths found through local search, perhaps through a less random mechanism than SARWs, may be more representative of information transmission in proteins than shortest paths (which still contain an element of randomness). Research needs to be done to know more about the communication pathways and information networks in protein molecules.

The positive picture painted by SARWs of PRNs is made possible in part by the high level of clustering or transitive interactions in PRNs. When the short-range links in PRNs which is the primary source of transitivity are randomized, i.e. in randSE networks, this positive picture deteriorates quite dramatically: walk lengths and mean path costs are no longer polynomially bounded but become linear with N (sections 3 & 6), communication becomes less specific (section 3) and more easily congested and fragile (sections 4 & 5). If the ability to interact or handle multiple stimulants simultaneously in a discriminatory way is important for protein function and stability, then from what we have observed in this research, a network topology with high clustering such as PRNs is now understandable.

**Acknowledgements**

This work was made possible by the facilities of the Shared Hierarchical Academic Research Computing Network (SHARCNET:www.sharcnet.ca) and Compute/Calcul Canada.